\documentclass[a4paper,12pt]{article}

\usepackage{graphicx,bm}
\usepackage[normalem]{ulem} % \sout{old text} for strikeout
\usepackage[dvipsnames]{xcolor} % For blue in-text comments and
\renewcommand\sout{\bgroup \color{red} \ULdepth=-.5ex \ULset}

\begin{document}

\begin{center}

{\large\bf
Decay angular distributions of $K^*$ and $D^*$ mesons  as a tool for the dynamics
of open strange and charm production}

\medskip

{Sang-Ho Kim$^{1}$, Yongseok Oh$^{2,1}$, and Alexander I. Titov$^{3\P}$ }

\medskip

$^1${Asia Pacific Center for Theoretical Physics, Pohang, Gyeongbuk 37673, Korea}\\
$^2${Department of Physics, Kyungpook National University, Daegu 41566, Korea}\\
$^3${Bogoliubov Laboratory of Theoretical Physics, JINR, Dubna 141980, Russia}\\
$^\P${E-mail: atitov@theor.jinr.ru}

\end{center}

\centerline{\bf Abstract}
We analyze decay spin-density matrix elements of $K^*$ and $D^*$ vector mesons
and find that the corresponding vector meson decay distributions
are sensitive to the production mechanisms.
This suggests that the measurement of these quantities can be used to unravel the production mechanisms. \\

%Keywords: Pion-baryon interactions, hadronic decays of mesons, quark-gluon string model \\

PACS:{ 13.85.-t, % 13.85.-t	Hadron-induced high- and super-high-energy interactions (energy > 10 GeV)
12.40.Nn,	% Regge theory, duality, absorptive/optical models (see also 11.55.Jy Regge formalism)
13.75.Gx,	% Pion-baryon interactions
13.25.-k	% Hadronic decays of mesons, hadronic decays of mesons
}\\

Investigation of open charm and open strangeness production processes is one of
the major hadron physics programs at current or planned accelerator facilities which are supposed to provide pion
beams~\cite{MNNS12} or antiproton beams~\cite{Friese06}.
These facilities are expected to produce high-quality beams at energies high enough to produce
strange or charm hadrons.

One of the key issues which are not fully resolved at present is the
charm/strangeness production mechanism in hadron reactions.
Since the reaction energy is not high enough to be treated asymptotically, the widely-used models
for heavy quark production based on perturbative QCD are not
applicable, and an essential improvement by including nonperturbative
contributions is indispensable.
Therefore, it is legitimate to rely on the approaches based on non-perturbative QCD
background for describing such peripheral reactions.
In the present work, we adopt the quark-gluon string model (QGSM) developed by Kaidalov and
collaborators in Refs.~\cite{Kaidalov80,Kaidalov82}, which has been applied for the evaluation
of cross sections of exclusive $\Lambda_c$ productions in $pp$ and
$\bar{p} p$ collisions~\cite{TK08,KKMW11} and in $\pi p$
collisions~\cite{BK83,KHKN15,KKH16,SOT}.
However, at relatively low energy the contribution of the pseudoscalar ($PS$) Reggeon exchange may be
comparable to that of the vector meson $(V)$ exchange.
Therefore, it is interesting and important to verify the similarities and differences of the PS and V exchanges in more detail
through various polarization observables.

In the present work, we elaborate on the angular distributions of pseudoscalar mesons originated from
the decays of vector mesons produced in $\pi N$ collisions.
More specifically, we consider the production of $K^*$ and $D^*$ vector mesons,
which decay into $K\pi$ and $D\pi$ final states, respectively.
Thus, the reaction under consideration in the present work is
$\pi^- + p \to V + Y \to (P + \pi) + Y$, where
$Y$, $V$, $P$ are flavored baryon, vector meson, and pseudoscalar meson, respectively.
In the strangeness sector, $Y = \Lambda(1116,1/2^+)$, $V = K^*(892,1^-)$,
and $P = K(494,0^-)$, while, in charm sector, $Y = \Lambda_c(2286,1/2^+)$, $V = D^*(2010,1^-)$,
and $P = D(1870,0^-)$~\cite{PDG16}.
The invariant amplitude of this reaction is expressed as
\begin{eqnarray}
T_{fi} = \mathcal{A}_{m_f^{}, \lambda_V^{}; \,m_i^{}}
%\frac{1}
\left[{p_V^2 - M_0^2 + i M_0 \Gamma_{\rm tot}}\right]^{-1}
\mathcal{D}_{\lambda_V^{}}(\Omega_f),
\label{EQ2}
\end{eqnarray}
where the amplitudes $\mathcal{A}$ and $\mathcal{D}$ determine the production and decay of
the intermediate vector mesons $V$ to the two pseudoscalars, respectively.
Here, $m_i^{}$ and $m_f^{}$ denote the spin projections of incoming and outgoing baryons,
respectively, and $\lambda_V^{}$ is the spin projection of the produced virtual vector meson.
$M_0$ and $\Gamma_{\rm tot}$ are the pole mass and the total decay width of the produced vector meson,
respectively.
The Mandelstam variables for the production process are defined as
$s = (p_\pi^{} + p_p^{})^2 = (p_V^{} + p_Y^{})^2$
and $t = (p_p^{} - p_Y^{})^2 = (p_\pi^{} - p_V^{})^2$, where $p_\pi^{}$, $p_p^{}$, $p_V^{}$, and
$p_Y^{}$ are the four momenta of the pion, proton, produced (virtual) vector meson, and hyperon,
respectively.
The solid angle and the magnitude of the three momentum of outgoing pseudoscalar mesons
in the rest frame of the vector meson are represented by $\Omega_f$ and $k_f$, respectively
The decay amplitude has a simple form of
$\mathcal{D}_\lambda = C \sqrt{{4\pi}/{3}} Y_{1\lambda}(\Omega_f)$,
where the constant $C$ is determined by the $V$-decay width and masses of
participating mesons. The details can be found in Ref.~\cite{SOT}.

The vector meson production amplitude determined within the QGSM or the standard Regge
model has the spin-independent form
for vector and pseudoscalar Reggeon exchanges as
${\cal A}^V\sim s^{\alpha^{V}(t)-1}$ and ${\cal A}^{PS}\sim s^{\alpha^{PS}(t)}$,
respectively. The unpolarized cross sections are discussed in many
works and can be found, for example, in Refs.~\cite{TK08,BK83} and will not be repeated here.

For studying polarization observables, we generalize the QGSM and standard Regge model for $V$- and $PS$-exchanges
by introducing spin-factors, which are uniquely defined by the symmetry considerations.
(For details, see, for example, Refs.~\cite{TK08,KHKN15,KKH16,SOT}.)
These factors read
\begin{eqnarray}
{\cal {M}}^V_{m_f^{},\lambda_V^{};m_i^{}} &=&
\epsilon^{\mu\nu\alpha\beta} q_{\mu}^{} p_{V\alpha}^{} \varepsilon^*_\beta (\lambda_V)
\times \bar u_{m_f^{}}(\Lambda)
\left[ (1+\kappa_{K^*p\Lambda}^{}) \gamma_\nu  - \kappa_{K^*p\Lambda}^{}
\frac{(p_p+p_\Lambda)_\nu}{M_p+M_\Lambda} \right] u_{m_i^{}}(p)/\mathcal{N}^V~,\nonumber\\
{\cal M}^{\rm PS}_{m_f^{},\lambda_V^{};m_i^{}} &=& \varepsilon^{*}_\mu(\lambda_V)\,q^\mu
\bar u_{m_f^{}}(\Lambda) \gamma_5 u_{m_i^{}}(p)/ \mathcal{N}^{PS},
\label{EQ18}
\end{eqnarray}
where $q = p_V^{} - p_\pi^{} = p_p^{} - p_\Lambda^{}$ is the momentum transfer and
$\kappa_{K^*p\Lambda}^{} (= 2.79)$~\cite{TK08} is the tensor coupling constant.
The Dirac spinors of initial baryon and final baryons are denoted by $u_{m_i^{}}$ and $u_{m_f^{}}$, respectively,
and $\varepsilon (\lambda_V^{})$ is the polarization vector of the produced vector meson.
Generalization to the case of charm production is achieved by the substitution
$M_\Lambda \to M_{\Lambda_c}$, $M_{K^*} \to M_{D^*}$, and so on.
Because of the lack of information, we assume $\kappa_{K^*p\Lambda}^{} = \kappa_{D^*p\Lambda_c}^{}$
as in Ref.~\cite{CERN-MPI-78}.
The normalization factor $\cal N$ in Eq.~(\ref{EQ18}) is introduced
to compensate the artificial $s$ and $t$ dependence generated by the spin factor.

The differential cross section is then obtained as
\begin{eqnarray}
\frac{d\sigma}{dt \, d\Omega_f} = \frac{d\sigma}{dt}\,W(\Omega_f),
\label{EQ10}
\end{eqnarray}
where
\begin{eqnarray}
W(\Omega_f) = \sum\limits_{m_i^{},m_f^{},\lambda_V^{},\lambda'_V}
\mathcal{M}_{m_f^{},\lambda_V^{};m_i^{}}
\mathcal{M}_{m_f^{},\lambda'_V;m_i^{}}^*
 \mbox{} \times
Y_{1\lambda_V^{}}(\Omega_f)\,Y^*_{1\lambda'_V}(\Omega_f).
\label{EQ11}
\end{eqnarray}
Here $ {d\sigma}/{dt}$ is the differential cross section of vector meson production,
and  $W(\Omega_f)$ is the decay angular distribution of outgoing pseudoscalar mesons.
For definiteness with isospin quantum number we consider
$K^{*0} \to K^+ \pi^-$ and $D^{*-} \to D^- \pi^0$ decays.
The decay angular distribution of outgoing $K^+$ depends on the choice
of quantization axis in the rest frame of the vector meson.
One may choose the quantization axis anti-parallel to the outgoing hyperon $Y$ in
the center-of-momentum frame of the production process.
Or the quantization axis may be defined to be parallel to the incoming pion, i.e., the
initial beam direction.
Following the convention of Ref.~\cite{CGLS72}, the former is called the $s$-frame and
the latter the $t$-frame.
The decay probabilities are expressed in terms of the spin-density matrix elements
$\rho_{\lambda\lambda^\prime}^{}$, where $\lambda_V^{}$ is abbreviated as $\lambda$,
which are determined by the amplitudes of Eq.~(\ref{EQ18}).
They are defined as
\begin{eqnarray}
\rho^0_{\lambda\lambda'} = \sum\limits_{m_i^{} = \pm\frac12,\, m_f^{} = \pm\frac12}
\mathcal{M}_{m_f^{},\lambda^{};\,m_i^{}}\, \mathcal{M}^*_{m_f^{},\lambda';\,m_i^{}} .
\label{EQ13}
\end{eqnarray}
More complicated cases with fixed polarization
of recoil baryon may be found in Ref.~\cite{SOT}.
Denoting the polar and the azimuthal angles of the outgoing pseudoscalar $K$ (or $D$) mesons
by $\Theta$ and $\Phi$, respectively, the decay angular distributions can be expressed
in terms of the spin-density matrix elements as
\begin{eqnarray}
W^{0}(\Omega_f) = \frac{3}{4\pi} \Bigl[ \rho^0_{00}\cos^2\Theta + \rho^0_{11}\sin^2\Theta
- \rho^0_{1-1} \sin^2\Theta \cos2\Phi
- {\sqrt{2}}\, \mathrm{Re}(\rho^0_{10}) \sin2\Theta\cos\Phi \Bigr].
\label{EQ15}
\end{eqnarray}

\begin{figure*}[t]
\centering
\includegraphics[width=0.7\textwidth]{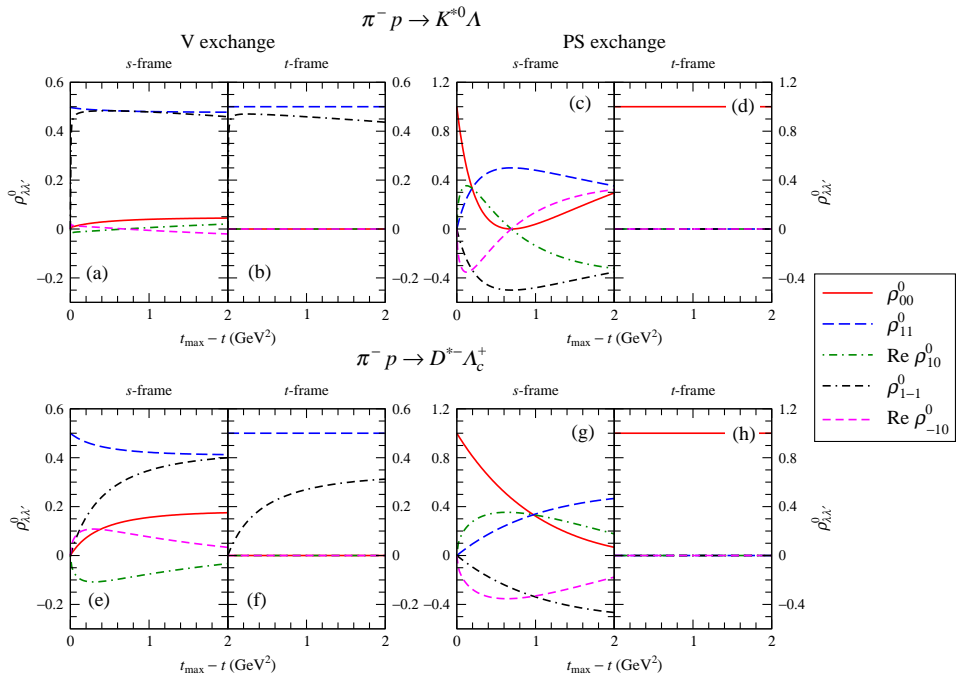}
\caption{The spin-density matrix elements $\rho^0_{\lambda\lambda'}$ defined in Eq.~(\ref{EQ13}) as
functions of $(t_{\rm max} - t)$. (a)--(d) for $K^{*0}$ production at
$p_\pi^{} = 6$~GeV/$c$ and (e)--(f) for $D^{*-}$
production at $p_\pi^{} = 15$~GeV/$c$.
The results for vector meson (V) and pseudoscalar (PS) Reggeon exchanges are in (a), (b), (e), (f)
and (c), (d), (g), (h) panels, respectively.
The results in (a), (c), (e), (g) are obtained in the $s$-frame, while those in (b), (d), (f), (h) are
in the $t$-frame.}
\label{Fig:4}
\end{figure*}

\begin{figure*}[h]
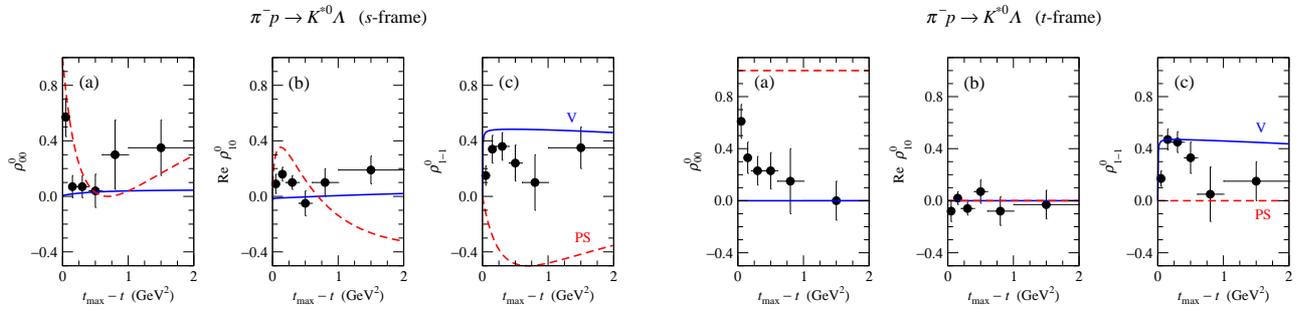

\centering
\includegraphics[width=0.45\textwidth]{Fig2.eps}\qquad
\includegraphics[width=0.45\textwidth]{Fig3.eps}
\caption{
Spin-density matrix elements for $K^{*0}$ production in the $s$- and $t$-frames.
The panels (a), (b), and (c) correspond to $\rho^0_{00}$, $\mbox{Re}\,\rho^0_{10}$, and $\rho^0_{1-1}$
matrix elements, respectively.
The vector and pseudoscalar Reggeon exchange models are depicted by the solid and dashed curves,
respectively.
The experimental data are from Ref.~\cite{CGLS72}.}
\label{Fig:6}
\end{figure*}

\begin{figure*}[t]
\centering
\includegraphics[width=0.7\textwidth]{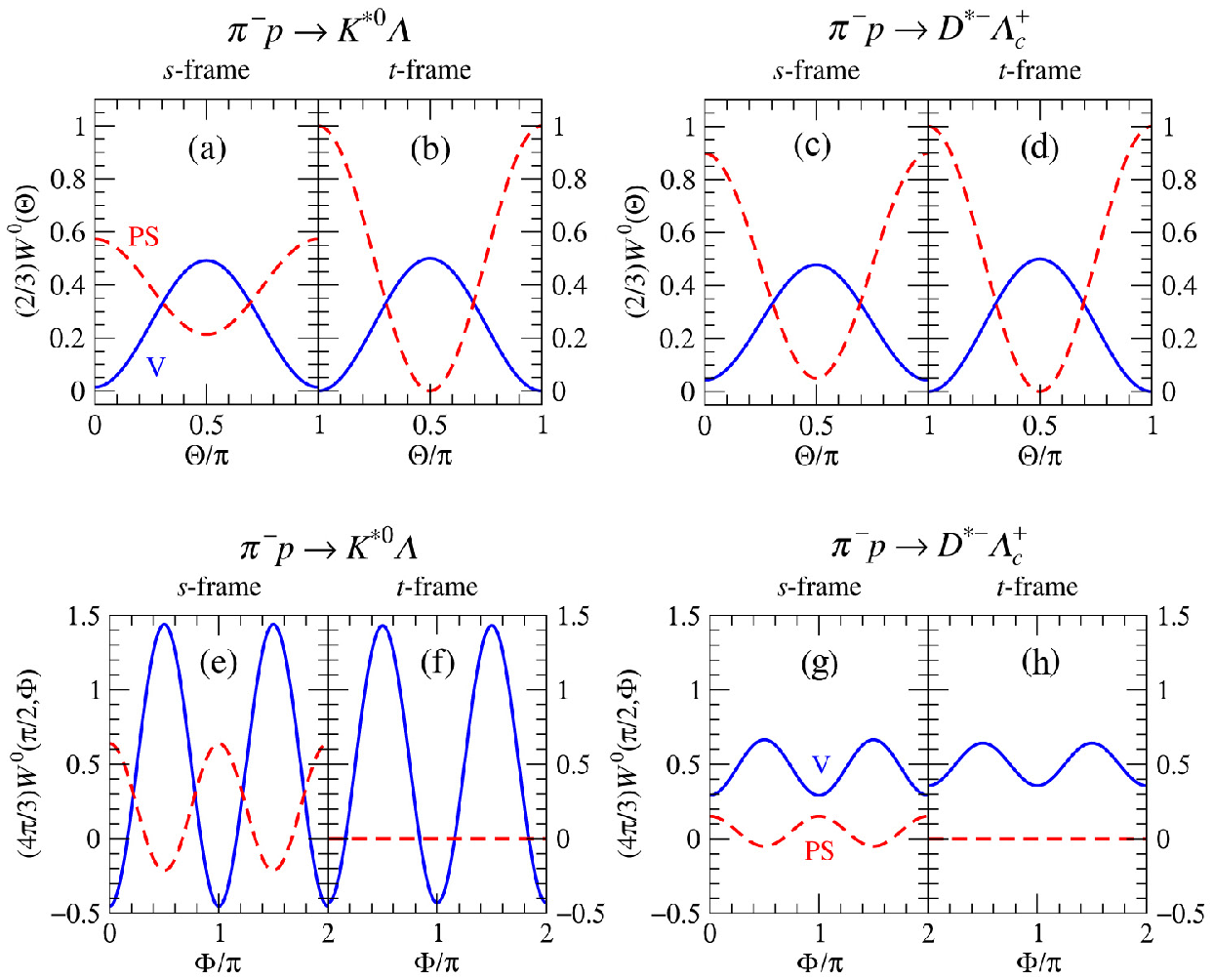}
\caption{
Angular distributions $\frac23\, W(\Theta)$ of Eq.~(\ref{EQ155}) for $K^{*0}$ and $D^{*-}$
are shown in (a,b) and (c,d), respectively.
Shown in (e,f) and (g,h) are the azimuthal angle distributions
$\frac{4\pi}{3}\,W(\Theta=\pi/2,\Phi)$ of Eq.~(\ref{EQ1555}).
The results are given at the $s$ frame (a,c,e,g) and the $t$ frame (b,d,f,h).
The vector and pseudoscalar Reggeon exchanges are depicted by the solid and dashed curves,
respectively.
Calculation is done for $|{t_{\rm max}-t}| = 0.1 \mbox{ GeV}^2$
at $p_\pi^{}$ = 6~GeV/$c$ for $K^{*0}$ production and 15~GeV/$c$ for $D^{*-}$ production.}
\label{Fig:8}
\end{figure*}

The results for the spin-density matrix elements $\rho^0_{\lambda\lambda'}$ defined in
Eq.~(\ref{EQ13}) are presented in Fig.~\ref{Fig:4} for $K^{*0}$  and $D^{*-}$ production
as functions of $(t_{\rm max} - t)$.
We also limit our consideration to relatively small values of $|{t}|$ such that $|{t_{\rm max}-t}| \leq 2$
GeV$^2$, where the applicability of the QGSM can be justified.
Shown in Fig.~\ref{Fig:4} are the results for the vector-type Reggeon exchange model and for
the pseudoscalar-type Reggeon exchange model, which are calculated in the $s$- and $t$-frames.
Our results numerically confirm the symmetry properties,
$\rho^0_{11} = \rho^0_{-1-1}$,  $\rho^0_{\pm1 0}=\rho^0_{0\pm 1}$,
$\rho^0_{\pm 10}=-\rho^0_{0\mp 1}$, and $\rho^0_{1-1}=\rho^0_{-11}$.

In the case of vector-type Reggeon exchange, the matrix elements $\rho^0_{\lambda\lambda'}$
with $|{\lambda}| = |{\lambda'}| = 1$ are enhanced.
This ascribes to the spin structure
$\epsilon^{\mu\nu\alpha\beta} q_{\mu}^{} p_{V\alpha}^{} \varepsilon^{*}_\beta(\lambda_V^{})$
of the amplitude shown in Eq.~(\ref{EQ18}).
In the vector meson rest frame, where $p_V^{} = (M_V^{},0,0,0)$ and $\bm{q} = - \bm{p}_\pi^{}$,
this factor is proportional to the vector product of ${\bm{\varepsilon}}^*(\lambda_V^{}) \times \bm{p}_\pi^{}$.
In the $s$-frame and small momentum transfers, $\bm{p}_\pi^{}$ has a large $z$ component
and a small $x$ component, which leads to ${\bm\varepsilon}^*(\lambda_V^{}) \times \bm{p}_\pi^{}
\simeq i\lambda_V {\bm\varepsilon}^*(\lambda_V^{})|\bm{p}_\pi|$
and thus causes the large enhancement of $\rho^0_{|{\lambda}| = 1,\, |{\lambda^\prime}| = 1}$.
In the $t$-frame, on the other hand, $\bm{p}_\pi^{}$ is parallel to the quantization axis, and
$\rho^0_{\lambda\lambda'}$ with either $\lambda = 0$ or $\lambda' = 0$ vanish.
We also note that $\rho_{1-1}^{0} = 0$ at $t = t_{\mathrm{max}}$, which arises
from the relation $\rho_{1-1}^{0} \propto \sin^2\theta$, where $\theta$ is the scattering
angle of the vector meson in the c.m. frame for the scattering process.
In the case of pseudoscalar-type Reggeon exchange, however, the situation is quite different.
The production amplitude of this mechanism is proportional to the scalar product,
$\bm{\varepsilon}^*(\lambda_V^{}) \cdot \bm{p}_\pi^{}$, which leads to a strong enhancement of
$\rho^0_{00}$ in the $t$-frame, so that $\rho^0_{00}=1$ and all the other
$\rho^0_{\lambda\lambda'}$ vanish.

In Fig.~\ref{Fig:6}, we compare our results with the available
experimental data of Ref.~\cite{CGLS72} for $K^{*0}$ production at $s$- and $t$-frames,
respectively.
Although the vector-exchange mechanism leads to a better agreement with the
data than the pseudoscalar-exchange model, we can see that the vector-exchange model
alone cannot successfully explain the data.
This conclusion is valid for comparison with data at
$p_{\pi}^{}=4.5$~GeV/$c$~\cite{CGLS72} as well.
 New experimental data for $K^*$ production with higher precision are, therefore,
strongly desired.
In $D^*$ production, the difference is also large enough to be verified by experiments and
the analyses can be done at current or planned experimental facilities.

The polar angle distributions of outgoing $K$ and $D$ mesons are obtained by integrating
$W(\Theta,\,\Phi)$ of Eq.~(\ref{EQ15}) over the azimuthal angle $\Phi$,
which gives
\begin{eqnarray}
\frac{2}{3}\,W^{0}(\Theta) &=& \rho^0_{00}\, \cos^2\Theta + \rho^0_{11}\, \sin^2\Theta .
\label{EQ155}
\end{eqnarray}
Correspondingly, the azimuthal angle distributions at a fixed polar angle $\Theta$ can also be obtained from
Eq.~(\ref{EQ15}).
At $\Theta=\frac{\pi}{2}$, we have
\begin{eqnarray}
\frac{4\pi}{3} W^{0}(\Theta=\frac{\pi}{2},\,\Phi) =
\rho^0_{11}-\rho^0_{1-1}\cos2\Phi .
\label{EQ1555}
\end{eqnarray}
These polar and azimuthal angle distributions are presented in Fig.~\ref{Fig:8}
in upper and bottom panels, respectively,
for the production and decays of
$K^*$ and $D^*$ mesons at $|{t_{\rm max}-t}| =  0.1 \mbox{ GeV}^2$ with $p_\pi^{} = 6$
and $15$~GeV/$c$, respectively. More details are given in figure captions.
In all cases, one can observe maxima at $\Theta=\frac{\pi}{2}$ for the vector trajectory exchange
while minima are observed at the same angle for the pseudoscalar trajectory exchange.
In other words, the distribution functions for the vector trajectory exchange display a cosine function
shape, while those for the pseudoscalar trajectory exchange show a sine function shape.
This is a direct consequence of the spin density matrix elements $\rho^0_{00}$ and
$\rho^0_{11}$ shown in Fig.~\ref{Fig:4}.

The corresponding azimuthal angle distributions are shown in the lower panels in Fig.~\ref{Fig:8}.
In the $s$-frame, the matrix element $\rho^0_{1-1}$ takes a positive value for vector-type
exchange and a negative value for pseudoscalar-type exchange.
This difference makes that $W(\frac{\pi}{2},\Phi)$ of vector-type exchange and
pseudoscalar-type exchange are out of phase.

In summary,  we investigated the reactions of open strangeness and open charm vector mesons
and found that contrary to the case of unpolarized cross section
the angular distributions of outgoing pseudoscalar meson is sensitive to the production
mechanism.
In particular, the measurements for $K^*$ and $D^*$ productions are complementary to each other
and would be important to understand the dependence of the production mechanisms on the
quark mass scale.
All these predictions can be tested and verified in future experimental programs with pion beams, for
instance, at J-PARC facility.

\section*{Acknowledgment}
The work of Y.O. was supported by the NRF of Korea under Grant No. NRF-2015R1D1A1A01059603.

\end{document}